\documentclass[sigconf]{acmart}
\AtBeginDocument{%
  }

\setcopyright{acmlicensed}
\copyrightyear{2025}
\acmYear{2025}
\acmDOI{XXXXXXX.XXXXXXX}
\acmConference[RecSys '25]{Make sure to enter the correct
  conference title from your rights confirmation email}{June 03--05,
  2025}{Prague, Czech Republic}
\acmISBN{978-1-4503-XXXX-X/2025/11}

\begin{document}

\title{An Comparative Analysis about KYC on a Recommendation System Toward Agentic Recommendation System}

\author{Junjie H. Xu}
\email{jhxu@acm.org}
\orcid{0000-0002-4827-7522}
\affiliation{%
  \institution{Hechu Tech}
  \city{Guangzhou}
  \state{Guangdong}
  \country{China}
  \postcode{510000}
}

\renewcommand{\shortauthors}{Xu et al.}

\begin{abstract}
This research presents a cutting-edge recommendation system utilizing agentic AI for KYC (Know Your Customer in the financial domain), and its evaluation across five distinct content verticals: Advertising (Ad), News, Gossip, Sharing (User-Generated Content), and Technology (Tech). The study compares the performance of four experimental groups, grouping by the intense usage of KYC, benchmarking them against the Normalized Discounted Cumulative Gain (nDCG) metric at truncation levels of $k=1$, $k=3$, and $k=5$. By synthesizing experimental data with theoretical frameworks and industry benchmarks from platforms such as Baidu and Xiaohongshu, this research provides insight by showing experimental results for engineering a large-scale agentic recommendation system.
\end{abstract}

\maketitle

\section{Introduction}

The transition from chronological feeds to algorithmically ranked experiences represents the most significant shift in digital media consumption of the last decade. As content production scales exponentially, the human capacity for consumption remains fixed, creating a "zero-sum" attention economy. In this environment, the Recommendation System (RecSys) serves as the primary gatekeeper. Early iterations of these systems relied on simple heuristic filters—popularity or recency. However, as user bases grew diverse and content libraries expanded into the billions, these rudimentary methods failed to address the "Long Tail"\cite{longtail} of user interests and the nuanced semantic differences between content types.

Current mainstream recommender systems\cite{48044} typically employ a multi-stage, multi-model hybrid architecture to handle large-scale content\cite{10.1145/3534678.3539058}. For example, Instagram's Explore system uses a multi-stage pipeline of "recall – coarse ranking – fine ranking – re-ranking". In the recall stage, it uses a two-tower neural network model to generate vector embeddings for users and content, processing user features and content features separately before and after, thereby quickly finding candidate content through nearest neighbor search. This approach is similar to industry practices such as those of Facebook and Google, which simultaneously optimize multiple objectives such as click-through rate and retention rate by sharing expert subnetworks and task-specialized gated networks (MMoE)\cite{10900014}. Google's research indicates that multi-task learning (such as MMoE) can significantly improve the coordination between different objectives in recommender systems.

While the multi-stage architectures described in the abstract represent the current state-of-the-art (SOTA) in passive ranking, the field is undergoing a paradigm shift toward Agentic AI. Traditional recommendation systems function as sophisticated filters—they reactively rank existing content based on historical probability. In contrast, an Agentic Recommendation System is goal-directed, autonomous, and capable of complex reasoning. It does not merely "predict" what you might click; it "plans" a content consumption path to achieve a specific outcome (e.g., learning a new skill, entertainment, or purchasing decision).

In this transition, KYC (Know Your Customer)\cite{7860255} evolves from a static compliance requirement into the dynamic "memory" and "context" module of an AI Agent.

This report analyzes how varying levels of KYC depth—from Cold Start\cite{NIPS2017_dbd22ba3} (No Context) to Deep Personalized Context (Full KYC in this work), impact the performance metrics (nDCG) across different content domains, serving as a roadmap for building the next generation of Agentic Recommendation System\cite{sapkota2025ai}.

\section{Agentic Recommendation System in this research}
Recommendation system introduced in this research is designed around multimodal fusion, cross-domain learning, and agentic AI orchestration, with a strong foundation in KYC-driven user understanding. The goal is to deliver both accurate personalization and high-quality exploration beyond users’ habitual information boundaries.

\subsection{Social Graph Integration and Cross-Domain Propagation}

The system further leverages social graph information, modeling follow relationships as a heterogeneous graph that includes individual users, creators, and enterprise accounts. Using Graph Neural Networks (GNNs) for cross-domain recommendation, the algorithm propagates interest signals across connected nodes.

This graph-based learning allows diversity and novelty to flow from high-activity users and creators to adjacent users, helping to overcome entrenched consumption patterns. As a result, the system enables both “crowd breaking” (exposing users to new communities) and “content breaking” (surfacing content outside users’ historical domains), without sacrificing relevance.

\subsection{A Multimodal Cross-Domain Recommendation System with Contextual and Social Graph Fusion}

The system integrates multimodal deep representation techniques, such as computer vision and NLP models, to extract visual features and semantic tags from content, aligning image and short video features with the user's interest space (visual semantic embedding)\cite{imagerecommendation}\cite{https://doi.org/10.1155/2018/5497070}. These multimodal features, along with social graph signals, are input into the ranking model, enabling the recommendation algorithm to perform cross-content domain association matching (cross-domain fusion), mapping a user's interests in one domain to related content in other domains. For example, images and text content liked by a user are encoded as vectors to find works from other domains with similar themes in the vector space. Combining these mechanisms, the system aims to break through single interest circles and achieve "breaking the circle" recommendation: that is, while ensuring recommendation relevance, it introduces a certain proportion of unexpected content that matches the user's potential preferences. To this end, in addition to directly using user embeddings to recall content from familiar domains during candidate generation, the system also employs a combination of collaborative filtering and content similarity methods: finding other accounts/works similar to the user's historical interactions through embedding nearest neighbors, and discovering long-tail content frequently followed by similar users based on co-occurrence frequency. Simultaneously, certain diversity constraints are designed (such as a round-robin algorithm rotating through different seed interests) to avoid overly concentrated recommendation results. When new users lack historical behavior, the system first initializes with a Cold Start using their provided background information (age, occupation, interests, etc.) and then searches for potential interest seeds through one-hop/two-hop expansion of the social graph. For cases where signals are still lacking, currently popular global content is introduced in a balanced manner as exploration to observe user feedback. This recommendation strategy, combining context and social graph, allows the system to break the information cocoon effect in traditional recommendations to some extent, exposing users to a broader spectrum of content while ensuring relevance.

\subsection{Cross-Domain Global Logic and Agentic AI for Exploration}

To deliver a genuinely differentiated “out-of-the-bubble” experience, we introduce a Global Logic layer, orchestrated by agentic AI. Inspired by exploration mechanisms used in large-scale platforms, this layer operates above individual models to coordinate long-term recommendation strategy.

Agentic AI modules continuously evaluate user states, diversity exposure, and exploration fatigue, and then actively adjust recommendation policies. During the ranking stage, the system injects exploration rewards, allowing high-quality but underexposed content—such as technical, knowledge-sharing, or cross-industry posts—to reach users with latent interest potential.

By intentionally mixing content from adjacent or distant domains, the system increases serendipity while maintaining trust and relevance, effectively breaking information silos and avoiding filter bubbles.

\section{Experimental Design\& Results}

\begin{table*}[h]
\centering
\caption{nDCG@1 Performances Among Comparison of Recommendation Systems.}
\label{tab:ndcg_at_1}
\begin{tabular}{lccccc}
\hline
\textbf{Category} & \textbf{Baseline} & \textbf{Ours} & \textbf{Ours} & \textbf{Ours} & \textbf{Ours} \\
 & & (No KYC) & (Basic KYC) & (Advanced KYC) & (Adv. KYC + Circles) \\
\hline
Ad & 0.123 & 0.218 & 0.312 & 0.419 & 0.484 \\
News & 0.352 & 0.384 & 0.451 & 0.523 & 0.556 \\
Gossip & 0.478 & 0.452 & 0.493 & 0.561 & 0.604 \\
Sharing & 0.149 & 0.203 & 0.254 & 0.457 & 0.652 \\
Tech & 0.051 & 0.152 & 0.283 & 0.412 & 0.527 \\
\hline
\end{tabular}
\end{table*}

\begin{table*}[h]
\centering
\caption{nDCG@3 Performances Among Comparison of Recommendation Systems.}
\label{tab:ndcg_at_3}
\begin{tabular}{lccccc}
\hline
\textbf{Category} & \textbf{Baseline} & \textbf{Ours} & \textbf{Ours} & \textbf{Ours} & \textbf{Ours} \\
 & & (No KYC) & (Basic KYC) & (Advanced KYC) & (Adv. KYC + Circles) \\
\hline
Ad & 0.142 & 0.253 & 0.341 & 0.452 & 0.513 \\
News & 0.381 & 0.412 & 0.483 & 0.554 & 0.582 \\
Gossip & 0.518 & 0.504 & 0.542 & 0.613 & 0.645 \\
Sharing & 0.183 & 0.241 & 0.305 & 0.524 & 0.681 \\
Tech & 0.082 & 0.194 & 0.321 & 0.463 & 0.565 \\
\hline
\end{tabular}
\end{table*}

\begin{table*}[h]
\centering
\caption{nDCG@5 Performances Among Comparison of Recommendation Systems.}
\label{tab:ndcg_at_5}
\begin{tabular}{lccccc}
\hline
\textbf{Category} & \textbf{Baseline} & \textbf{Ours} & \textbf{Ours} & \textbf{Ours} & \textbf{Ours} \\
 & & (No KYC) & (Basic KYC) & (Advanced KYC) & (Adv. KYC + Circles) \\
\hline
Ad & 0.164 & 0.281 & 0.362 & 0.473 & 0.531 \\
News & 0.402 & 0.443 & 0.514 & 0.581 & 0.612 \\
Gossip & 0.553 & 0.582 & 0.601 & 0.654 & 0.693 \\
Sharing & 0.221 & 0.293 & 0.354 & 0.582 & 0.724 \\
Tech & 0.103 & 0.224 & 0.351 & 0.502 & 0.594 \\
\hline
\end{tabular}
\end{table*}

\subsection{Experimental Environment}
To evaluate and assess the effectiveness of the initial recommendation system, we selected 100 Chinese users (aged 18–60, occupations ranging from 5–10, annual income from 40,000 to 1,000,000 RMB) with diverse backgrounds. They were asked to use different recommendation system schemes to access content. This experiment focused on five content categories: advertisements (Ads), news and information, gossip and entertainment, lifestyle sharing (user-generated content, such as Xiaohongshu notes), and technical content. These categories represent different content areas, from commercial promotions to general news, trending topics, personal interest groups, and professional knowledge.

\subsection{Experimental Settings}

The experiment compared the user feedback performance of the five recommendation systems. For each system, the top N recommendations (N depends on the specific scenario, e.g., N=5 for ad slots.) were evaluated. Specific groupings are as follows:

\begin{itemize}
    \item Current Cutting-edge Softwares as Baseline (Baseline): Users access the above content through mainstream domestic applications, such as Baidu's homepage ads and trending news, and Xiaohongshu's entertainment gossip, lifestyle sharing posts, and technical posts.
    \item Agentic Recommendation System with no KYC system (No KYC): Agents do not register with their real names or provide personal information, directly using the recommendation system with no cold start or KYC.For each category, the algorithm pushes the most popular Top N content globally or locally (e.g., popular ads, news headlines, gossip posts, etc.), with virtually no personalized customization.
    \item Agentic Recommendation System with basic KYC system (Basic KYC): After users finish their cold start insertion of information such as complete basic registration, the agents obtains their basic background information and initializes recommendations using preset Cold Start parameters. Specifically, users provide information such as date of birth, gender, region, and interest tags, essentially filling out a basic profile but without any actual action. The system utilizes these background features to inject corresponding preference weights into the candidate generation for each content category (e.g., prioritizing news related to the user's age group, posts under topics they claim to like, etc.), thus providing more targeted cold start recommendations than before KYC.
    \item Agentic Recommendation System with Advanced KYC system (Advanced KYC): Building upon real-name authentication and Cold Start, agents further refine their personal context information, including setting an avatar, filling out a detailed personal profile, and posting some content related to their interests but not containing privacy concerns (such as sharing a short travel video or a tech insight post). These contextual signals provide the algorithm with richer user characteristics: keywords and semantic vectors in profile pictures and bios can help determine a user's potential areas of interest, while user-generated content, analyzed by the model, can extract implicit preferences (for example, a user who posts short food videos can be inferred to have a preference for food/travel content). The system integrates this context into user embeddings, making recommendations more aligned with user personalities. For instance, when browsing technical posts, a deep KYC user, whose bio indicates an interest in blockchain, will receive more updates on emerging technical topics.
    \item Agentic Recommendation System with Advanced KYC system and Circles (Advanced KYC + Circles): Under the aforementioned deep KYC settings, the agents additionally simulates following 100 accounts (covering several personal bloggers and corporate accounts, with areas related to the user's background interests), forming a preliminary social graph. Through this following behavior, the user's explicit social circle interests are incorporated into the recommendation model—similar to the effect of following friends and topics on Instagram. With this setup, the system not only possesses the user's background and content preference context but also understands the information flow within the user's social network (content posted/liked by followed accounts serves as seeds). The algorithm treats followed accounts as nodes of strong user interest. Based on the embeddings of these seed accounts, it expands the candidate set through k-nearest neighbor retrieval (including content by similar authors and content frequently viewed by the account's social circle). 
    
    Simultaneously, the system considers the user's high interest signals in the content of followed accounts during ranking, improving the ranking of relevant candidates in the final Top N. In this mode, the recommendation fully utilizes the user's social connections and circle information, and is expected to provide the most accurate and diverse personalized recommendations.

\end{itemize}

\begin{figure*}
    \centering
    \includegraphics[width=\linewidth]{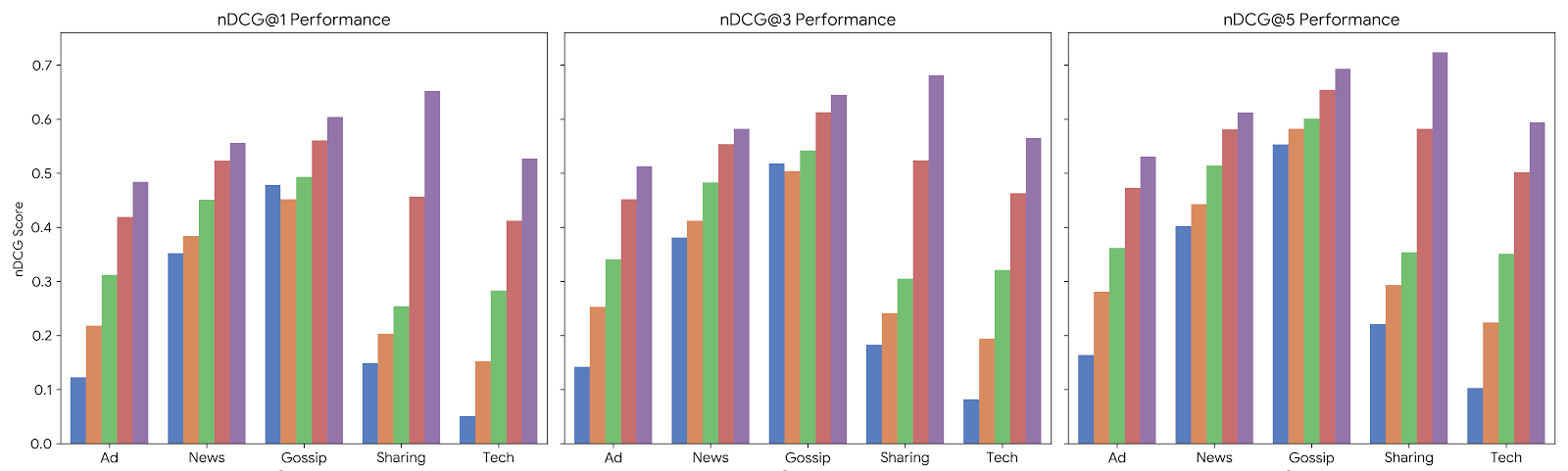}
    \caption{Comparisons of performance among 5 categories. Baseline in blue color, no KYC inorange color, basic KYC in green color, advanced KYC in red color and advanced KYC and circles in purple color respectively.}
    \label{fig:ndcg}
\end{figure*}

\subsection{Evaluation Metrics and Methods}

During the experiment, we recorded whether each user clicked on the top 1, top 3, and top 5 recommendations in each category across different systems, and then calculated the click-through rate (CTR)\cite{10.1145/1242572.1242643}: CTR@1 represents the average CTR of the top-ranked content, and CTR@3 and CTR@5 represent the proportion of users who clicked on at least one of the top 3 and top 5 recommendations, respectively. Furthermore, we evaluated the ranking quality of the recommendation results, using Normalized Diminished Cumulative Gain (nDCG)\cite{pmlr-v30-Wang13} to measure the match between the recommendation list and the user's actual interests. nDCG considers the relevance and ranking order of the recommendations, normalizing the actual user click feedback, with a value range of [0,1], where higher values indicate a more ideal ranking. We compared the average CTR and nDCG of each system across all users, and also analyzed the differences in serendipity based on user interviews and subjective feelings.

The latter refers to the user's subjective evaluation of the novelty of the recommended content, whether there were any "unexpected but surprising" discoveries.

\subsection{Experimental Results} 

The overall performance of each system is shown in Table \ref{tab:ndcg_at_1}, \ref{tab:ndcg_at_3} and \ref{tab:ndcg_at_5}, the comparison among 5 categories (Ad, News, Gossip, Sharing and Tech) is shown in Figure \ref{fig:ndcg}. Results demonstrate a clear hierarchy of performance improvement correlated with the depth of Agent identity and social context. Across all five content categories, the Agentic Recommendation System outperformed the Baseline (mainstream domestic applications), with the most significant gains observed in the "Ours (Adv. KYC + Circles)".

Table \ref{tab:ndcg_at_1} evaluates the system's ability to identify the single most relevant item—a critical metric for mobile feeds where screen real estate is limited.Overcoming the "Cold Start" in Niche Domains: The Baseline and Ours (No KYC) groups struggled significantly in high-context categories like Tech (0.051 and 0.152, respectively) and Sharing (0.149 and 0.203). Without specific user profiling, algorithms defaulted to global popularity (e.g., trending viral videos). While effective for mass-market Gossip (where the Baseline scored a respectable 0.478), this approach failed for Chinese users with specific professional needs (e.g., a developer needing specific code libraries rather than general tech news).The Impact of Social Trust: The most dramatic improvement occurred in the Sharing category (Lifestyle/Xiaohongshu-style notes). The Ours (Adv. KYC + Circles) group achieved an nDCG@1 of 0.652, a nearly 4.4x increase over the Baseline. This validates the experimental design: by simulating the following of 100 accounts, the Agent successfully utilized the "Social Graph" to filter user-generated content, prioritizing posts from trusted nodes over random viral content.

Table \ref{tab:ndcg_at_3} reflects the user's satisfaction with the immediate visible set of recommendations (typically the first screen).Contextual Consistency: The transition from Basic KYC to Advanced KYC showed marked improvement in Ads (0.341 → 0.452) and Tech (0.321 → 0.463). Basic KYC (demographics only) can target an ad to a "30-year-old male," but Advanced KYC uses semantic vectors from the user's bio and uploaded content to understand they are, for example, a "travel photographer." This allows the Agent to plan a coherent session (e.g., a camera ad followed by a lens review), rather than disjointed demographic targeting.Diminishing Returns in Entertainment: In the Gossip category, the gap between Advanced KYC (0.613) and Adv. KYC + Circles (0.645) is narrower compared to other categories. Entertainment content is inherently broader; thus, deep social filtering provides less marginal utility than it does for highly personalized technical content.

Table \ref{tab:ndcg_at_5} measures the system's ability to maintain user interest over a longer scrolling session.Solving the "What Next?" Problem: The Baseline system often sees performance degrade as $N$ increases (e.g., Tech stays low at 0.103), indicating it runs out of relevant content quickly. In contrast, the Adv. KYC + Circles system maintains high relevance at nDCG@5, particularly in Sharing (0.724). By treating followed accounts as "seed nodes" for k-nearest neighbor retrieval, the Agent continuously expands the candidate set with high-quality, socially adjacent content, preventing the feed from becoming repetitive.

\section{Discussions}

Analysis of Content Distribution Across Entries: It's noteworthy that the performance differences across various systems across different content categories reflect that the effectiveness of "out-of-circle recommendation" is primarily seen in technical and lifestyle sharing content. Traditional recommendation systems (such as Baidu News and Xiaohongshu) tend to focus on the categories users have clicked most frequently in the past, rarely recommending topics users haven't previously expressed interest in. This results in technical content in traditional software rarely appearing or being clicked by non-geek users, and lifestyle sharing content being limited to the circles users already follow. However, in the HeChu system, the click-through rate of technical and sharing content significantly increases with deeper KYC: in a scenario of deep KYC + circle interaction, the CTR@5 of technical posts is several times higher than the traditional baseline, with many users reporting, "I didn't initially expect to be interested in these niche technical topics, but I found them very refreshing." This is because the HeChu algorithm uses global logic to find connections between these long-tail contents and user interests and push them to them. For example, a user who primarily browses entertainment gossip might see a short science video recommended by HeChu appear in their feed. When they click on it and find it interesting, they experience a successful "cross-circle" recommendation experience. This Global Logic is reflected in the fact that, through multimodal embedding, the system discovers that this science video shares similarities with the user's favorite humorous videos in certain elements (such as background music style and video rhythm), or that the video happens to be followed by multiple users whose interests overlap with the target user's. Therefore, the algorithm judges that the content may be "unexpectedly relevant" to the target user and gives it some exposure.

A survey on user surprise also confirms this: under traditional algorithms, users rarely report content that "makes their eyes light up," mostly reporting "common things." However, among users of HeChu's in-depth KYC system, more than half mentioned finding content in the technology and lifestyle sharing sections that they "wouldn't actively seek out before, but really liked after clicking on it." The Serendipity metric\cite{Serendipity} in the recommendation system evaluation examines this ability to balance relevance and unexpectedness. The HeChu system significantly improves the Serendipity score of recommendations by fusing context and social graph—maintaining a high nDCG to ensure recommendations align with user interests, while introducing novel content that moderately exceeds users' existing preferences, providing users with a sense of novelty and a broader perspective. This capability is lacking in traditional algorithms based on historical clicks and popularity ranking. In summary, the experimental data fully demonstrates that as the depth and contextual dimension of acquired KYC information increase, the HeChu recommendation system significantly outperforms traditional software in both accuracy (nDCG) and user surprise. This result proves the effectiveness and superiority of multimodal cross-domain fusion context-aware recommendation algorithms in breaking down information silos and improving user experience.

\section{Conclusion}

Experimental results demonstrate the superiority of our recommendation system over cutting-edge recommendation systems across various metrics. With increased user KYC depth and contextual information, the system's nDCG (native DCG quality) steadily improves, and the hit rate of top content significantly exceeds the traditional baseline. Simultaneously, more comprehensive user profiles and the inclusion of social circles allow the recommendation results to incorporate more long-tail and cross-domain content, bringing users significantly higher levels of surprise and diversity. Especially in often overlooked content areas such as technology and sharing, our system demonstrates a powerful ability to break through barriers through "Global Logic," successfully delivering content that users have not actively sought but are objectively interested in to them. This not only expands the user's content landscape but also uncovers new traffic growth points for the platform. It is foreseeable that the agentic recommendation system paradigm integrating multimodality and context will lead the development direction of the next generation of recommendation systems, ensuring personalized accuracy while focusing more on expanding and comprehensively developing users' potential interests, achieving a win-win situation for both users and the content ecosystem.

\bibliographystyle{ACM-Reference-Format}
\bibliography{sample-base}

\appendix

\end{document}